\documentclass[aps,prd,superscriptaddress,showpacs,preprintnumbers]{revtex4}
\usepackage{graphicx,color}
\usepackage{bm}
\newcommand{\be}{\begin{equation}}
\newcommand{\ee}{\end{equation}}
\newcommand{\bea}{\begin{eqnarray}}
\newcommand{\eea}{\end{eqnarray}}

\newcommand{\bfk}{\mbox{\boldmath $k$}}

\def\kt{k_\perp}
\newcommand{\bfp}{\mbox{\boldmath $p$}}

\newcommand{\bfP}{\mbox{\boldmath $P$}}
\newcommand{\bfS}{\mbox{\boldmath $S$}}

\newcommand{\pup}{p^\uparrow}

\def\lsim{\mathrel{\rlap{\lower4pt\hbox{\hskip1pt$\sim$}}\raise1pt\hbox{$<$}}}
\def\gsim{\mathrel{\rlap{\lower4pt\hbox{\hskip1pt$\sim$}}\raise1pt\hbox{$>$}}}
\def\nostrocostruttino#1\over#2{\mathrel{\mathop{\kern 0pt \rlap
{\hbox{$#1$}}} \hbox{\kern-.135em $#2$}}}

\begin{document}
%%%%%%%%%%%%%%%%%%%%%%%%%%%%%%%%%%%%%%%%%%%%%%%%%%%%%%%%%%%%%%%%%%%%%%%%%%%%%%

\title{Sivers effect and the single spin asymmetry $A_N$ in
$\pup p \to h \, X$ processes}

\author{M.~Anselmino}
\affiliation{Dipartimento di Fisica Teorica, Universit\`a di Torino,
             Via P.~Giuria 1, I-10125 Torino, Italy}
\affiliation{INFN, Sezione di Torino, Via P.~Giuria 1, I-10125 Torino, Italy}
\author{M.~Boglione}
\affiliation{Dipartimento di Fisica Teorica, Universit\`a di Torino,
             Via P.~Giuria 1, I-10125 Torino, Italy}
\affiliation{INFN, Sezione di Torino, Via P.~Giuria 1, I-10125 Torino, Italy}
\author{U.~D'Alesio}
\affiliation{Dipartimento di Fisica, Universit\`a di Cagliari, Cittadella
             Universitaria, I-09042 Monserrato (CA), Italy}
\affiliation{INFN, Sezione di Cagliari,
             C.P.~170, I-09042 Monserrato (CA), Italy}
\author{S.~Melis}
\affiliation{Dipartimento di Fisica Teorica, Universit\`a di Torino,
             Via P.~Giuria 1, I-10125 Torino, Italy}
\affiliation{INFN, Sezione di Torino, Via P.~Giuria 1, I-10125 Torino, Italy}
\author{F.~Murgia}
\affiliation{INFN, Sezione di Cagliari, C.P.~170, I-09042 Monserrato (CA), Italy}
\author{A.~Prokudin}
\affiliation{Jefferson Laboratory, 12000 Jefferson Avenue, Newport News, VA 23606, USA}
\date{\today}

\begin{abstract}
  The single spin asymmetry $A_N$, for large $P_T$ single inclusive particle production in $\pup p$ collisions, is considered within a generalised parton model and a transverse momentum dependent factorisation scheme. The focus is on the Sivers effect and the study of its potential contribution to $A_N$, based on a careful analysis of the Sivers functions extracted from azimuthal asymmetries in semi-inclusive deep inelastic scattering processes. It is found that such Sivers functions could explain most features of the $A_N$ data, including some recent STAR results which show the persistence of a non zero $A_N$ up to surprisingly large $P_T$ values.
\end{abstract}


\maketitle

\section{\label{1}Introduction}

Among the leading-twist Transverse Momentum Dependent Partonic Distribution Functions (TMD-PDFs, often shortly referred to as TMDs), the Sivers distribution~\cite{Sivers:1989cc,Sivers:1990fh,Boer:1997nt} is most interesting and widely investigated. It describes the number density of unpolarised quarks $q$ (or gluons) with intrinsic transverse momentum $\bfk _\perp$ inside a transversely polarised proton $\pup$, with three-momentum $\bfP$ and spin polarisation vector $\bfS$,
\be
\hat f_ {q/\pup} (x,\bfk_\perp) = f_ {q/p} (x,\kt) +
\frac{1}{2} \, \Delta^N \! f_ {q/\pup}(x,\kt) \;
{\bfS} \cdot (\hat {\bfP}  \times \hat{\bfk}_\perp)
\,,\label{sivnoi}
\ee
where $x$ is the proton light-cone momentum fraction carried by the quark, $f_ {q/p}(x,\kt)$ is the unpolarised TMD ($\kt = |\bfk_\perp|$) and $\Delta^N \! f_ {q/\pup}(x,\kt)$ is the Sivers function. $\hat {\bfP} = \bfP/|\bfP|$ and $\hat{\bfk}_\perp = \bfk_\perp/\kt$ are unit vectors.
Notice that the Sivers function is most often denoted as $f_{1T}^{\perp q}(x, k_\perp)$~\cite{Mulders:1995dh}; this notation is related to ours by~\cite{Bacchetta:2004jz}
\be
\Delta^N \! f_ {q/\pup}(x,k_\perp) = - \frac{2\,k_\perp}{m_p} \>
f_{1T}^{\perp q}(x, k_\perp) \>. \label{rel}
\ee

A knowledge of the Sivers distribution allows a modelling of the 3-dimensional momentum structure of the nucleon~\cite{Boer:2011fh} and, possibly, an estimate of the parton orbital angular momentum~\cite{Bacchetta:2011gx}.

All the available information on the Sivers function has been obtained from SIDIS data, $\ell \, N \to \ell \, h \, X$, and the study of the azimuthal distribution of the final hadron $h$ around the $\gamma^*$ direction in the $\gamma^* - N$ centre of mass (c.m.) frame. This analysis is based on the TMD factorisation scheme~\cite{Ji:2004xq,Ji:2004wu,Bacchetta:2008xw,Collins:2011book}, according to which the SIDIS cross section is written as a convolution of TMD-PDFs, Transverse Momentum Dependent Fragmentation Functions (TMD-FFs) and known elementary interactions. Such a scheme holds in the kinematical region defined by
\be
P_T \simeq k_\perp \simeq \Lambda_{\rm QCD} \ll Q \>,
\ee
where $P_T$ is the magnitude of the final hadron transverse momentum. The presence of the two scales, small $P_T$ and large $Q$, allows to identify the contribution from the unintegrated partonic distributions ($P_T \simeq \kt$), while remaining in the region of validity of the QCD
parton model. The study of the QCD evolution of the Sivers and unpolarised TMDs -- the so-called TMD evolution -- has much progressed lately~\cite{Ji:2004xq,Ji:2004wu,Collins:2011book,Aybat:2011zv,Aybat:2011ge,Echevarria:2012pw, Echevarria:2012qe}, with the first phenomenological applications~\cite{Aybat:2011ta,Anselmino:2012aa,Bacchetta:2013pqa, Godbole:2013bca,Sun:2013dya,Boer:2013zca,Sun:2013hua}.

The extraction of the Sivers functions from SIDIS data can then be performed on a sound ground. This has been done for the first time in Refs.~\cite{Anselmino:2005ea,Anselmino:2008sga,Vogelsang:2005cs,
Collins:2005ie,Anselmino:2005an}, exploiting HERMES~\cite{Diefenthaler:2007rj} and COMPASS~\cite{Martin:2007au} data, and resulting in a reasonable knowledge of the Sivers functions for $u$ and $d$ quarks, although in a limited range of $x$ values, $x \lesssim 0.3$.

Much literature has emphasised the special interest and the peculiar properties of the Sivers effect. Trying to understand its origin at the partonic level has related the possibility of a non zero Sivers function with final~\cite{Brodsky:2002cx} or initial~\cite{Brodsky:2002rv} state interactions, respectively in SIDIS and Drell-Yan (D-Y) processes. This, in turns, induces a process dependence of the effect itself. The most clear-cut consequence is the prediction of an opposite sign of the Sivers functions when contributing to single spin asymmetries (SSAs) in SIDIS and D-Y processes~\cite{Collins:2002kn}; as polarised D-Y experiments have never been performed so far, such a prediction has not been tested yet. Crucial information might be available in the future from $\pup p$ experiments at RHIC, Fermilab or from the COMPASS hadronic run at CERN, with pions colliding on a polarised nucleon target. The TMD factorisation scheme, valid for SIDIS processes, holds for D-Y as well, where the small and large scale are respectively the total transverse momentum ($q_T$) and the invariant mass ($M$) of the leptonic pair.

In this paper we focus on another class of puzzling results which strongly challenge our understanding of high energy strong interactions, that is the SSAs, usually denoted by $A_N$, measured in $\pup p \to h \, X$ inclusive reactions and defined as:
\be
A_N = \frac{d\sigma^\uparrow - d\sigma^\downarrow}
           {d\sigma^\uparrow + d\sigma^\downarrow}
\quad\quad {\rm with} \quad\quad
d\sigma^{\uparrow, \downarrow} \equiv
\frac{E_h \, d\sigma^{p^{\uparrow, \downarrow} \, p \to h \, X}}
{d^{3} \bfp_h} \>, \label{an}
\ee
and where $\uparrow, \downarrow$ are opposite spin orientations perpendicular to the scattering plane, in the $\pup p$ c.m.~frame. $A_N$ differs from the SSAs of SIDIS and D-Y processes because in such a case there is only one large scale in the process -- the transverse momentum $P_T$ of the final observed hadron -- and there is no small scale related to the intrinsic motions, both in the distribution and fragmentation functions, which are integrated over. The TMD factorisation scheme used for SIDIS and D-Y processes has not been proven in this case.

Large values of $A_N$ have been measured since a long time in many different experiments. The first ones were at relatively low energy~\cite{Klem:1976ui,Krueger:1998hz,Allgower:2002qi,Antille:1980th,
Adams:1991rw,Adams:1991cs,Adams:1991rv,Adams:1991ru}, and the common expectation was that such asymmetries would vanish at higher energies; however, data from RHIC at $\sqrt s = 62.4$~\cite{Arsene:2008aa}, 200~\cite{Adams:2003fx,Adler:2005in,Lee:2007zzh,Abelev:2008af,Adamczyk:2012xd}
or even 500~\cite{Igo:2012,Bland:2013pkt} GeV, still show puzzling non zero values of $A_N$.

Several approaches to understanding $A_N$, within QCD and some sort of factorisation scheme, can be found in the literature. All of them, directly or indirectly, are related to the Sivers function or other TMDs.

A QCD collinear factorisation formalism at next-to-leading-power (twist-3) has been developed and used in the phenomenological studies of $A_N$~\cite{Efremov:1981sh,Efremov:1984ip,Qiu:1991pp,Qiu:1998ia,
Kanazawa:2000hz,Kanazawa:2000kp,Kouvaris:2006zy,Kanazawa:2010au,Kanazawa:2011bg}.
In this approach the spin effect is not embedded in a spin dependent TMD, but the necessary phase for generating the non-vanishing SSAs arises from the quantum interference between an elementary scattering amplitude with one active collinear parton and an amplitude with two active collinear partons.
The SSAs are therefore proportional to some non-probabilistic three-parton correlation functions, which are convoluted with product of amplitudes, rather than cross sections. These amplitudes are process dependent, while the three-parton correlation functions are universal.

However, one can show that the twist-3 three-parton correlation functions have a close connection with the $\kt$-moment of the TMD-PDFs; in particular the quark-gluon correlator is related to the first $\kt$-moment of the SIDIS Sivers function~\cite{Boer:2003cm}. It has been recently pointed
out~\cite{Kang:2011hk} that the quark-gluon correlation functions, as obtained from the Sivers functions extracted from SIDIS data~\cite{Anselmino:2005ea,Anselmino:2008sga}, indeed lead to sizeable values of $A_N$, which agree in magnitude with the measured ones, but {\it with the wrong sign} (the so-called sign mismatch problem). A recent analysis~\cite{Gamberg:2013kla} of the spin asymmetry $A_N$ for single inclusive jet production in $\pup p$ collisions collected by the A$_N$DY experiment~\cite{Nogach:2012sh} does not show the same sign problem; however, the measured asymmetry is very small.

An alternative, more phenomenological approach, is based on the assumption of the validity of the TMD factorisation also for $\pup p \to h \, X$ processes~\cite{Sivers:1989cc,Sivers:1990fh,Anselmino:1994tv,Anselmino:1998yz,Anselmino:1999pw, D'Alesio:2004up,Anselmino:2005sh,D'Alesio:2007jt}; it generalises the usual collinear factorisation scheme (Generalised Parton Model, GPM) and the single inclusive cross section is written as a convolution of TMD-PDFs, TMD-FFs and QCD partonic cross sections. In that it adopts the same scheme which holds for SIDIS and D-Y processes with one small and one large scale. In this model the spin effects are included in the TMDs, which are supposed to be process-independent.

More recently, a third approach has been proposed~\cite{Gamberg:2010tj, Ratcliffe:2009sz}, which assumes the TMD factorisation as in the GPM, but takes into account and absorbs the initial and final state interactions, {\it i.e.} the process dependence of the Sivers function, in the elementary interactions. In such a scheme the cross section is a convolution of process-independent TMDs with process-dependent hard parts; these modified hard parts are very similar in form to those in the
twist-3 collinear approach. It turns out that this modified GPM formalism leads to results and predictions opposite to those of the conventional GPM~\cite{Gamberg:2010tj}.

In this paper we explore the possibility of understanding the experimental results on $A_N$ in $\pup p \to h \, X$ processes with the Sivers effect and within the generalised parton model of Refs.~\cite{D'Alesio:2004up,Anselmino:2005sh,D'Alesio:2007jt}. The first phenomenological applications
of the Sivers effect~\cite{Anselmino:1994tv,Anselmino:1998yz,D'Alesio:2004up} in hadronic interactions considered the Sivers function as a free input, not constrained by SIDIS data.
In Ref.~\cite{Boglione:2007dm} it was shown that the use of the Sivers functions, as extracted from SIDIS data, could in principle explain the SSAs observed at RHIC, both in size and in sign. We further pursue this study, with a careful analysis of the SIDIS extracted Sivers functions, with their
uncertainties, and investigate whether such functions, assumed to be process independent, can explain the data on $A_N$, including the most recent ones. A similar study has been recently completed~\cite{Anselmino:2012rq} regarding the Collins effect~\cite{Collins:1992kk}, with the conclusion that it cannot, alone, explain all the available data on $A_N$.

\section{Sivers effect and $A_N$ in the Generalised Parton Model formalism}

The Generalised Parton Model~\cite{D'Alesio:2004up,Anselmino:2005sh,D'Alesio:2007jt,Anselmino:2012rq}, can be considered as a natural phenomenological
extension of the usual collinear factorisation scheme, with the inclusion of
spin and $k_\perp$ effects through the TMDs and the dependence of the elementary interactions on the parton intrinsic motions; it was actually first proposed,
for unpolarised processes, in Ref.~\cite{Feynman:1978dt}. In this approach the
single spin effect, $d\sigma^\uparrow \not= d\sigma^\downarrow$, originates from
the TMDs; in Ref.~\cite{ Anselmino:2004ky} and its
correction~\cite{Anselmino:2012rq} it was shown that the only non negligible contributions to $A_N$ are given by the Sivers TMD-PDF and the Collins TMD-FF,
\be
A_N = \frac{[d\sigma^\uparrow - d\sigma^\downarrow]_{\rm Sivers}
+ [d\sigma^\uparrow - d\sigma^\downarrow]_{\rm Collins}}
{d\sigma^\uparrow + d\sigma^\downarrow} \>\cdot \label{ansc}
\ee
The Collins contribution was studied in Ref.~\cite{Anselmino:2012rq}, while this paper is devoted to the Sivers effect.

In our GPM scheme the contribution of the Sivers effect to the numerator of $A_N$, for $\pup p \to h \, X$ large $P_T$ processes, is given by
\bea
[d\sigma^\uparrow - d\sigma^\downarrow]_{\rm Sivers}
&=& \!\!\! \sum_{a,b,c,d} \int \frac{dx_a \, dx_b \, dz}
{16 \, \pi^2 \, x_a \, x_b \, z^2 s} \; d^2 \bfk_{\perp a} \,
d^2 \bfk_{\perp b}\, d^3 \bfp_{\perp}\,
\delta(\bfp_\perp \cdot \hat{\bfp}_c) \> J(p_{\perp}) \>
\delta(\hat s + \hat t + \hat u) \nonumber\\
&\times& \Delta^N\!f_{a/\pup}(x_a, k_{\perp a}) \,
\cos (\phi_a)\,
 f_{b/p}(x_b, k_{\perp b}) \> \frac{1}{2}
 \left[ |\hat M_1^0|^2 + |\hat M_2^0|^2 + |\hat M_3^0|^2 \right]_{ab\to cd} \>
D_{h/c}(z, p_\perp) \>, \label{numans}
\eea
where $\Delta^N\!f_{a/\pup}(x_a,k_{\perp a})$ is the Sivers function for parton $a$, Eqs.~(\ref{sivnoi}, \ref{rel}), which couples to the unpolarised TMD for parton $b$, $f_{b/p}(x_b, k_{\perp b})$, and the unpolarised fragmentation function $D_{h/c}(z, p_\perp)$ of parton $c$ into the
final observed hadron $h$. $\bfp_\perp$ is the transverse momentum of hadron $h$ with respect to the 3-momentum $\bfp_c$ of its parent fragmenting parton. $J(p_{\perp})$ is a kinematical factor, which at
${\cal O}(p_\perp/E_{h})$ equals 1. For details and a full explanation of the notations we refer to Ref.~\cite{Anselmino:2005sh} (where $\bfp_\perp$ is denoted as $\bfk_{\perp C}$).

The phase factor $\cos(\phi_a)$ originates directly from the $\bfk_\perp$ dependence of the Sivers distribution [${\bfS} \cdot (\hat {\bfP} \times \hat{\bfk}_\perp)$, Eq.~(\ref{sivnoi})], while the $\hat M_i^0$ are the three independent hard scattering helicity amplitudes defined in
Ref.~\cite{Anselmino:2005sh}, describing the lowest order QCD interactions.
The sum of their moduli squared is proportional to the elementary unpolarised cross section $d\hat\sigma^{a b \to c d}$, that is
\be
\label{eq:sigma}
\frac{d\hat\sigma^{ab\to cd}}{d\hat t} = \frac{1}{16\pi\hat s^2}\,\frac{1}{2} \sum_{i=1}^3 |\hat M_i^0|^2\,.
\ee
The explicit expressions of $\sum_i|\hat M_i^0|^2$, which give the QCD dynamics in Eq.~(\ref{numans}), can be found, for all possible elementary interactions, in Ref.~\cite{Anselmino:2005sh}. The QCD scale is chosen as $Q = P_T$.

The denominator of Eq.~(\ref{an}) or (\ref{ansc}) is twice the unpolarised cross section and is given in our TMD factorisation by the same expression as in Eq.~(\ref{numans}), where one simply replaces the factor $\Delta^N\!f_{a/\pup}\,\cos(\phi_a)$ with $2f_{a/p}$.
In Ref.~\cite{Boglione:2007dm} it was shown that such an expression leads to results for the unpolarised cross section in agreement with data.

We can now use the information so far available on the Sivers functions as extracted from SIDIS data and give some realistic estimates for the Sivers contribution to $A_N$ for several single-inclusive large $P_T$ particle production in $\pup p$ collisions. More specifically, we will consider the Sivers effect for inclusive pion, kaon, photon and jet production and will see how much it can contribute to the available experimental data on $A_N$. The analogue of Eq.~(\ref{numans}) for direct photon and inclusive jet production will be given below.

\subsection{The Sivers functions in SIDIS and $\pup p \to h\, X$ processes
\label{sec:scan}}

Let us start by considering the available information on the Sivers functions and
the procedure followed to obtain them. The first extraction -- from now on denoted
as SIDIS-1 fit -- was presented in Ref.~\cite{Anselmino:2005ea}, where the MRST01
set for the unpolarised PDFs~\cite{Martin:2002dr} and the Kretzer set for the
unpolarised FFs~\cite{Kretzer:2000yf} were adopted. An updated extraction of the
Sivers functions -- SIDIS-2 fit -- was presented in Ref.~\cite{Anselmino:2008sga}.
In this case, the GRV98 set for the unpolarised PDFs~\cite{Gluck:1998xa} and the
pion and kaon FFs by de Florian, Sassot and Stratmann (DSS)~\cite{deFlorian:2007aj} were considered. Notice that the use of different PDFs does not make any relevant difference; therefore, in the following, we will consider only the GRV98 set.

The main features of the parameterisations adopted in those studies are the following: the analysis of SIDIS data is performed at leading order, ${\cal O}(k_\perp/Q)$, within the proven TMD factorisation approach for SIDIS, where $Q$ is the large scale in the process. A simple factorised
form of the TMD functions was adopted, using a Gaussian shape for their $k_\perp$ dependent component. For the unpolarised parton distribution and fragmentation functions we have:

\be
f_{q/p}(x,k_\perp) = f_{q/p}(x)\,
\frac{e^{-k_\perp^2/\langle k_\perp^2 \rangle}}
{\pi \langle k_\perp^2 \rangle}
\quad\quad\quad
D_{h/q}(z,p_\perp) = D_{h/q}(z)\,
\frac{e^{-p_\perp^2/\langle p_\perp^2 \rangle}}
{\pi \langle p_\perp^2 \rangle}\,,
\label{eq:pdf-ff-unp}
\ee
where $\langle k_\perp^2\rangle$ and  $\langle p_\perp^2 \rangle$ have been fixed by analysing the Cahn effect in unpolarised SIDIS processes, see Ref.~\cite{Anselmino:2005nn}:
\be
\langle k_\perp^2\rangle = 0.25\, {\rm GeV}^2\,, \qquad\qquad
\langle p_\perp^2\rangle = 0.20\, {\rm GeV}^2\,.
\label{eq:k-p-cahn}
\ee
The recently introduced TMD evolution was not taken into account, while we considered the DGLAP QCD evolution of the collinear factorised part.

The Sivers functions, $\Delta^N f_{q/p^\uparrow}(x,k_\perp)$, have been parameterised as follows:
\be
\Delta^N\! f_{q/p^\uparrow}(x,k_\perp) = 2 \, {\cal N}_q^S(x)\,f_{q/p}(x)\,
h(k_\perp)\,\frac{e^{-k_\perp^2/\langle k_\perp^2 \rangle}}
{\pi \langle k_\perp^2 \rangle}\,,
\label{eq:siv-par}
\ee
where
\be
{\cal N}_q^S(x) = N_q^S x^{\alpha_q}(1-x)^{\beta_q}\,
\frac{(\alpha_q+\beta_q)^{(\alpha_q+\beta_q)}}
{\alpha_q^{\alpha_q}\beta_q^{\beta_q}}\,,
\label{eq:nq-coll}
\ee
with $|N_q^{S}|\leq 1$, and
\be
h(k_\perp) = \sqrt{2e}\,\frac{k_\perp}{M}\,e^{-k_\perp^2/M^2}\,.
\label{eq:h-siv}
\ee

With these choices, the Sivers functions automatically fulfil their proper positivity bounds for any $(x,k_\perp)$ values. For the $Q^2$ evolution of the Sivers function, as commented above, we consider the unpolarised DGLAP evolution of its collinear factor $f_{q/p}(x)$. Notice that in the SIDIS-1 fit we actually exploited also a different (power-like) functional form for $h(k_\perp)$, still controlled by a single parameter, leading to almost no differences in our results. In what follows we will only use the functional form given in Eq.~(\ref{eq:h-siv}).

In order to reduce the number of free parameters, some additional assumptions were adopted. Concerning the SIDIS-1 fit, we considered only $u$ and $d$ quarks Sivers functions, with flavour dependent $\alpha$ and $\beta$ parameters. This amounts to a total of 7 parameters:
\be
N_u, \,N_d,\, \alpha_u,\, \alpha_d,\, \beta_u,\, \beta_d,\, M\,.
\label{eq:7-par}
\ee

In the SIDIS-2 fit, since we were aiming also at explaining some large kaon SIDIS azimuthal asymmetries, we tentatively included also the Sivers functions for antiquarks and strange quarks, $\bar u$, $\bar d$, $s$ and $\bar s$. To keep the number of parameters under control we then assumed flavour independent $\alpha$ and $\beta$ parameters for the sea quarks ($\alpha_{\rm sea}, \beta_{\rm sea}$). Moreover, since the large $x$ behaviour of the Sivers function could not, and still cannot, be constrained by SIDIS data (see a more detailed comment below), we also assumed a single flavour independent $\beta$ parameter, equal for quarks and antiquarks. This amounts to a total of 11 free parameters:
\be
N_u,\, N_d,\, N_{\bar u},\, N_{\bar d},\, N_s,\, N_{\bar s},\, \alpha_u,\, \alpha_d,\, \alpha_{\rm sea},\, \beta, M\,.
\label{eq:11-par}
\ee
Notice that even with such a choice, our complete parameterisation of the Sivers functions, Eq.~(\ref{eq:siv-par}), allows for further differences among parton flavours, which are contained in the usual unpolarised PDFs.

Both fits gave good results. Nevertheless it is worth stressing the main differences in the two extractions, which indeed play an important role in the present study. In fact, a direct use of SIDIS-1 results in the computation of SSAs in $\pup p\to h \, X$ processes for RHIC kinematics, as presented in Ref.~\cite{Boglione:2007dm}, gave very encouraging results.
Notice that at that time the Collins effect was believed to be suppressed~\cite{ Anselmino:2004ky}. On the other hand, if we use the SIDIS-2 fit to compute the same SSAs we would get too small $A_N$ values, the reason being the different $\beta$ values coming from the two fits.

More generally, as discussed in the context of the Collins SIDIS azimuthal asymmetries for the transversity distributions~\cite{Anselmino:2012rq}, a study of the statistical uncertainties of the best fit parameters clearly shows that SIDIS data are not presently able to constrain the large $x$
behaviour of the quark ($u$, $d$) Sivers distributions, leaving a large uncertainty in the possible values of the parameter $\beta$. This is due to the limited range of Bjorken $x$ values currently explored by HERMES and COMPASS experiments, $x_B \lesssim 0.3$. In this respect the large $x_B$ results expected from JLab 12 GeV experiments will be precious~\cite{Gao:2010av,Gao:2011zzb}.

This uncertainty plays a crucial role when one tries to study the SSAs in hadronic collisions starting from the results obtained from SIDIS data, because the largest pion SSAs are measured at large Feynman
$x$ values, $x_F \gsim 0.3$, which implies $x\gsim 0.3$.

To investigate the role of the Sivers effect in explaining the large value of $A_N$ in $\pup p$ collisions we should therefore carefully explore the large $x$ behaviour of the Sivers functions.
To this aim we follow the same strategy we have devised in the recent study of the contribution of the Collins effect to $A_N$, the so-called ``scan procedure''~\cite{Anselmino:2012rq}. Here we summarise schematically its main steps and motivations.
\begin{itemize}
\item
  The $\beta_q$ parameters, which control the large $x$ behaviour of the TMDs and are largely undetermined by SIDIS data, play instead an important role in the computation of $A_N$, which is sizeable mainly in the large $x_F$ region. We can notice this explicitly by comparing, as commented above, the different implications on $A_N$ of the SIDIS-1 and the SIDIS-2 fits. In our choice of the independent parameters it is then natural to allow for a flavour dependence of $\beta$, limited, because of the relevance of the large $x$ region, to the valence quark contributions. More explicitly, we only use the PDFs for $u$ and $d$ valence quarks in the Sivers functions~(\ref{eq:siv-par}) and the contribution of sea quarks and gluons is neglected in the sum over partons in Eq.~(\ref{numans}).
\item
  We start the scan procedure by performing a preliminary 7-parameter [those of Eq.~(\ref{eq:7-par})] ``reference fit" to SIDIS data. This reference best fit will have a total $\chi^2 = \chi^2_0$. We then let the two parameters $\beta_u$ and $\beta_d$ vary, choosing them in the range $0.0$---$4.0$ by discrete steps of $0.5$, and for each of the 81 pairs of fixed $\beta$s we perform a new 5-parameter fit to SIDIS data.
\item
  As a next step we select only those fits leading to a $\chi^2$ such that $\chi^2 \leq \chi^2_0 + \Delta\chi^2$. Notice that, since the reference fit and the scan fits have a different number of free parameters, the selection criterion is applied to the total $\chi^2$ rather than to the $\chi^2$ per degree of freedom, $\chi^2_{\rm dof}$. The chosen value of $\Delta\chi^2$ is the same as that used to generate the error band, following the procedure described in the Appendix A of Ref.~\cite{Anselmino:2008sga}. We find (for 217 data points) $\chi^2_0 = 270.51$ and $\Delta\chi^2 = 14.34$. As expected from the arguments given above, all 81 fits lead to acceptable $\chi^2$ values for SIDIS data; this further confirms the observation that the SIDIS data are not sensitive to the large $x$ behaviour of the Sivers function.
\item
  We then compute, for each of the 81 selected sets, the contribution of the Sivers effect to $A_N$, according to Eqs.~(\ref{an})--(\ref{numans}). We do that for pion and kaon production in the kinematical regions of the STAR and BRAHMS experiments at RHIC. The corresponding results span the shaded areas (scan band) which are shown in the figures of our results. When compared with the experimental available data, the scan bands show the potentiality of the Sivers effect alone to account for the measured values of $A_N$ in $p^\uparrow p\to h\,X$ processes, while preserving a fair description (quantified by $\Delta\chi^2)$ of the SIDIS data on the Sivers azimuthal asymmetry.
\item
  We have considered in our scan procedure all available SIDIS
data~\cite{Airapetian:2009ae,Alekseev:2008aa}, with the exception of the recent
ones by the COMPASS Collaboration off a transversely polarised proton
target~\cite{Adolph:2012sn}. As shown in Refs.~\cite{Aybat:2011ta,Anselmino:2012aa}
the analysis of these data, reaching higher $Q^2$ values, requires a careful use of
the proper TMD evolution, which is ignored here, as a correct implementation of the
TMD evolution in $\pup p \to h \, X$ large $P_T$ processes is so far unknown. We have
checked that the $\chi^2_{\rm dof}$ of our fits would be approximately (30-40)\% worse
for the SIDIS data including the proton COMPASS results and no TMD evolution.
\item
  We study the contribution of the Sivers effect to the SSA $A_N$ at RHIC energies only, although it might contribute also to the (larger) SSAs measured at lower energies~\cite{Klem:1976ui,Krueger:1998hz,Allgower:2002qi,Antille:1980th, Adams:1991rw,Adams:1991cs,Adams:1991rv,Adams:1991ru}. The reason is that we consider only the processes for which our GPM and TMD factorisation can reasonably well reproduce the unpolarised cross section~\cite{Boglione:2007dm}.
\end{itemize}

\subsubsection{Results from the scan procedure}

\begin{figure}[t]
\begin{minipage}[c]{19cm}
\includegraphics[width=14cm,angle=0]{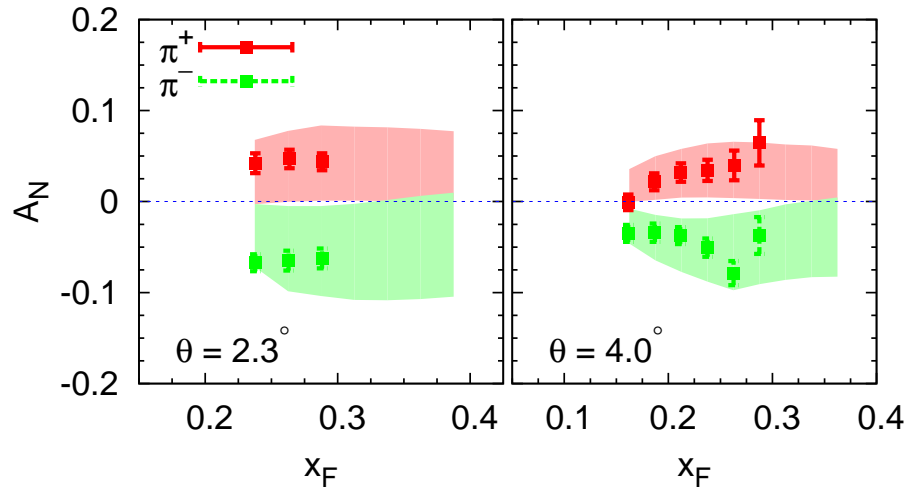}
\end{minipage}
\caption{
  Scan band ({\it i.e.} the envelope of the 81 curves obtained with the scanning procedure) for the Sivers contribution to the charged pion single spin asymmetries $A_N$, at $\sqrt s =$ 200 GeV, as a function of $x_F$ at two different scattering angles, compared with the corresponding BRAHMS experimental data~\cite{Lee:2007zzh}. The shaded scan band is generated, adopting the GRV98 set of collinear PDFs and the Kretzer FFs, following the procedure explained in the text.
}
\label{fig:an-brahms-pi}
\end{figure}

\begin{figure}
\begin{minipage}[c]{19cm}
\includegraphics[width=14cm,angle=0]{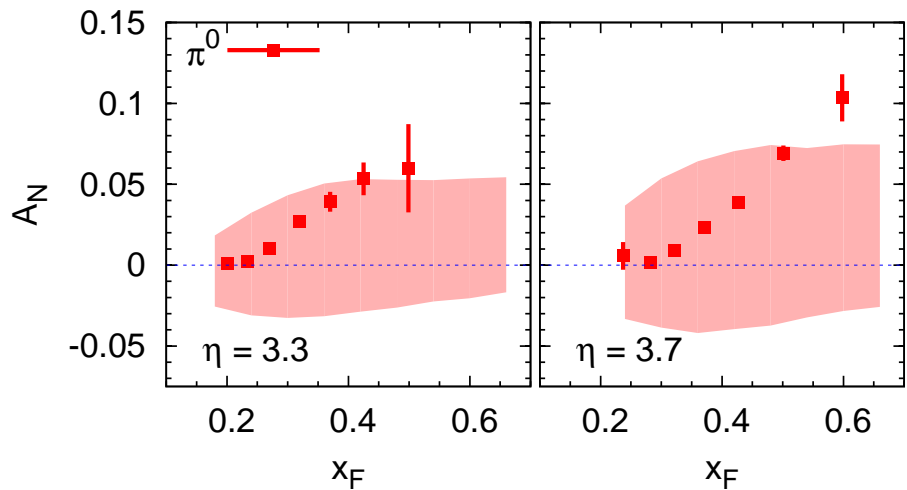}
\end{minipage}
\caption{
  Scan band ({\it i.e.} the envelope of the 81 curves obtained with the scanning procedure) for the Sivers contribution to the neutral pion single spin asymmetry $A_N$, at $\sqrt s =$ 200 GeV, as a function of $x_F$ at two different pseudo-rapidity values, compared with the corresponding STAR experimental data~\cite{Abelev:2008af}. The shaded scan band is generated, adopting the GRV98 set of collinear PDFs and the Kretzer FFs, following the procedure explained in the text.
}
\label{fig:an-star}
\end{figure}

\begin{figure}
%\begin{minipage}[c]{19cm}
\includegraphics[width=14cm,angle=0]{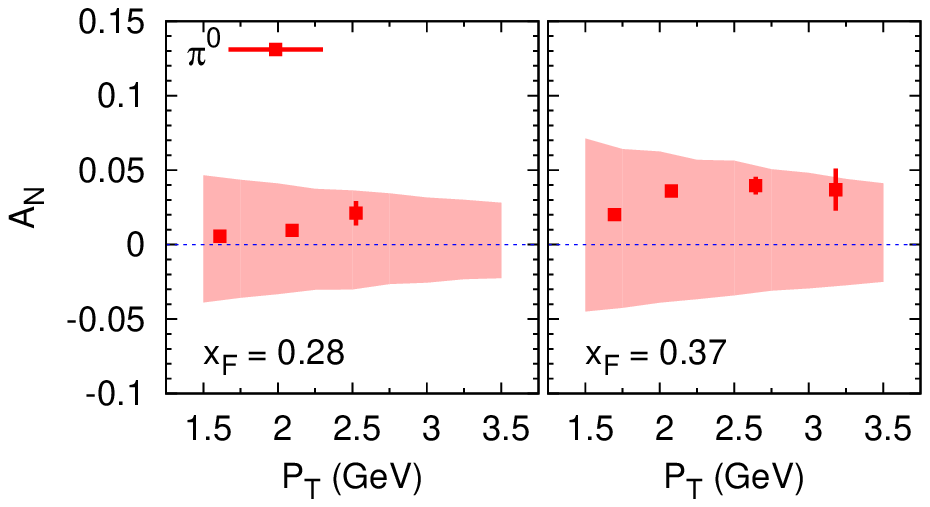}
\\
%\end{minipage}
\includegraphics[width=14cm,angle=0]{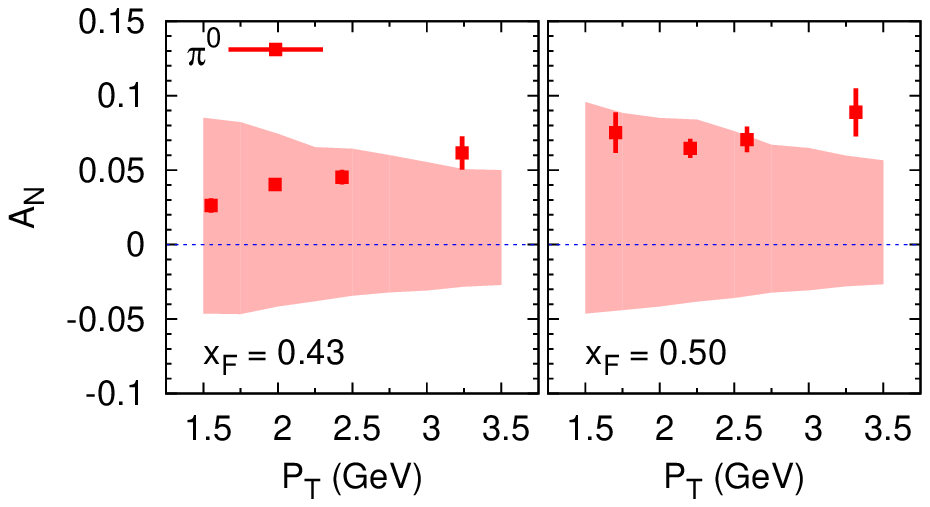}
\caption{
  The same as in Fig.~\ref{fig:an-star}, but with the STAR data plotted vs.~the pion transverse momentum, $P_T$, for different bins in $x_F$, $\langle x_F \rangle =$ 0.28, 0.37, 0.43 and 0.50. } \label{fig:an-star-xf}
\end{figure}

\begin{figure}
%\begin{minipage}[c]{19cm}
\includegraphics[width=8cm,angle=0]{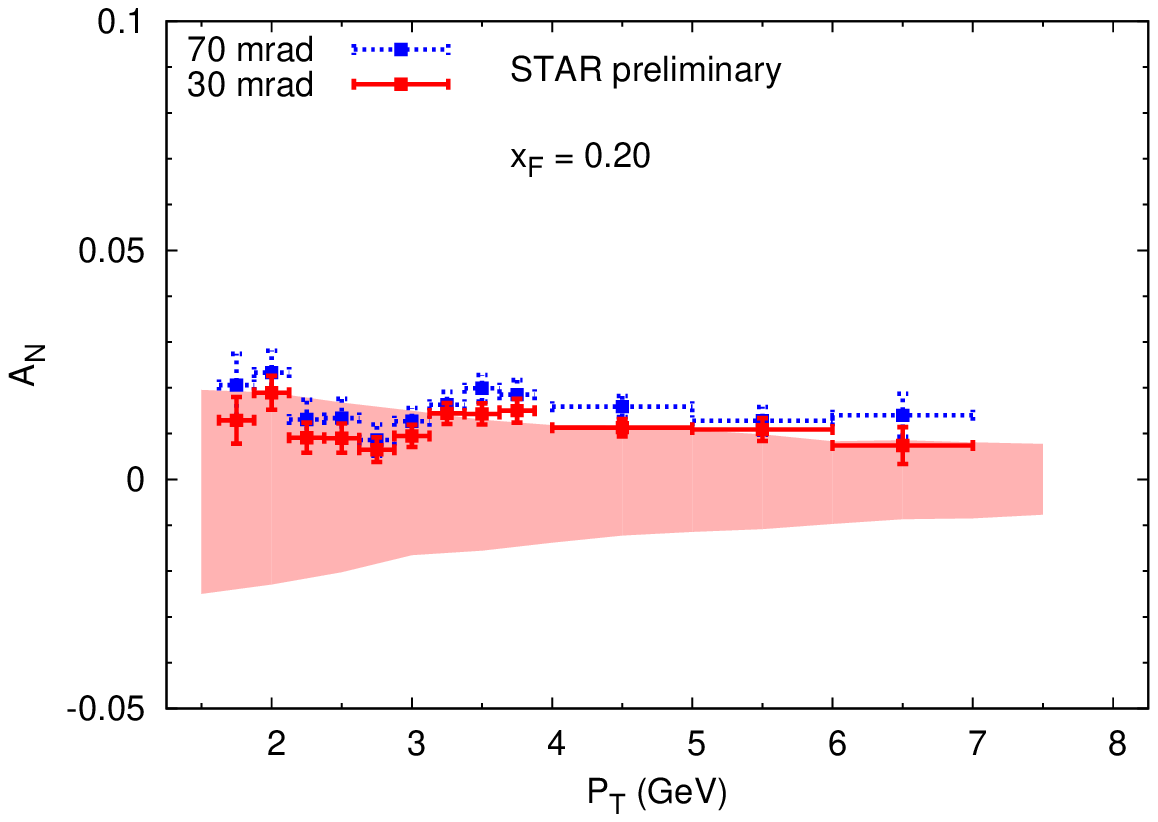}
%\end{minipage}
\includegraphics[width=8cm,angle=0]{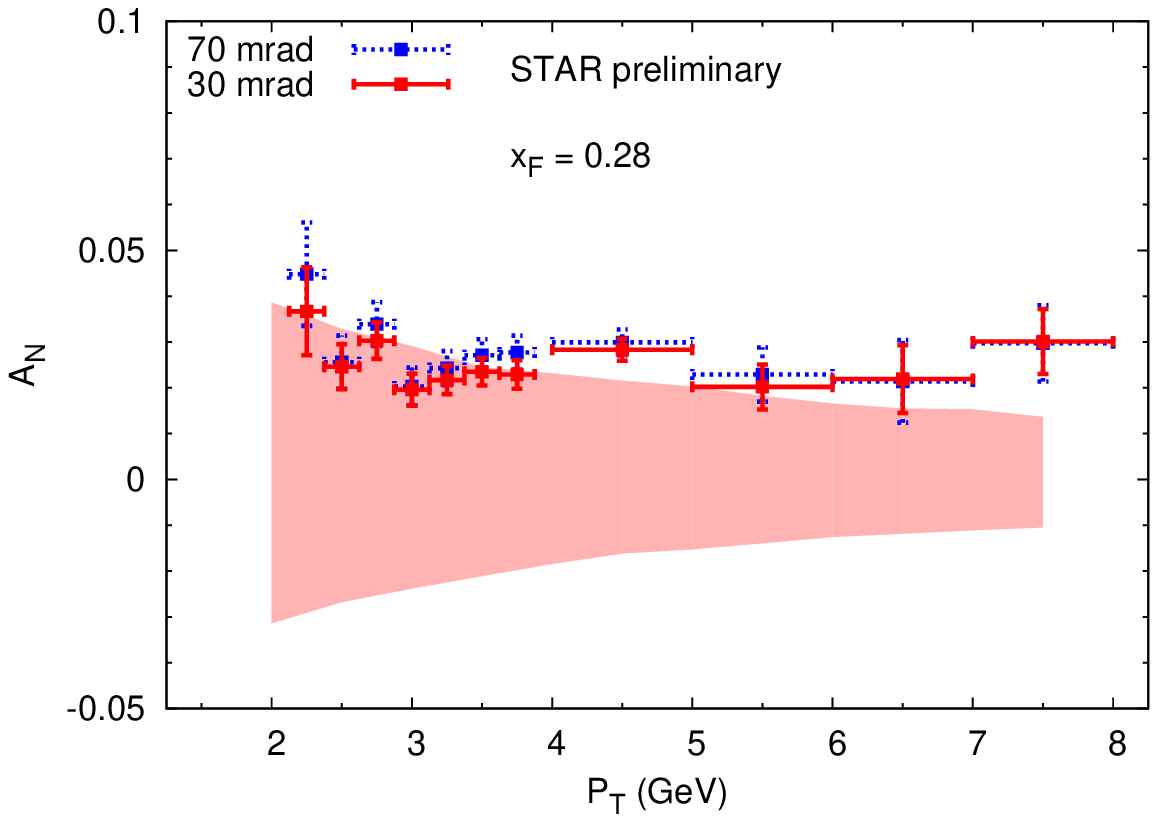}
\vskip -12pt
\caption{
  Scan band ({\it i.e.} the envelope of the 81 curves obtained with the scanning procedure) for the Sivers contribution to the neutral pion single spin asymmetry $A_N$, as a function of $P_T$ for different $x_F$ values at $\sqrt s = 500$ GeV, compared with the corresponding STAR preliminary experimental data at $\langle x_F \rangle = 0.20, 0.28$~\cite{Igo:2012}. The shaded scan band is generated, adopting the GRV98 set of collinear PDFs and the Kretzer FF set, following the procedure explained in the text.}
\label{fig:an-star-500}
\end{figure}

Some of our results for RHIC experiments are shown in Figs.~\ref{fig:an-brahms-pi}-\ref{fig:an-star-500}. We have computed $A_N$ adopting, as explained above, a single set of collinear parton distributions~\cite{Gluck:1998xa} and two different sets for the pion and kaon collinear FFs~\cite{Kretzer:2000yf,deFlorian:2007aj}; the results shown correspond to the Kretzer set. Other results not shown are very similar and would not add any significant information.

%%%%%%%%%%%%%% Fig. 1
%
%%%%%%%%%%%%%% Fig. 2

%%%%%%%%%%%%%% Fig. 3

Let us start by considering the case of inclusive pion production. This will also help a direct comparison with the corresponding study on the potential role of the Collins contribution to the same observable~\cite{Anselmino:2012rq}.

In Fig.~\ref{fig:an-brahms-pi} the scan band for $A_N$, as a function of $x_F$ at fixed scattering angles, is shown for charged pions and BRAHMS kinematics, while in Fig.~\ref{fig:an-star} the same result is given, at fixed pseudo-rapidity values, for neutral pions and STAR kinematics. We also give the scan band, as a function of $P_T$ at several fixed $x_F$ values, for STAR kinematics in Fig.~\ref{fig:an-star-xf}. All these results are given at $\sqrt s = 200$ GeV. We then consider the latest and interesting preliminary data obtained by STAR at large $P_T$ and $\sqrt s = 500$ GeV~\cite{Igo:2012}, and show our scan band in Fig.~\ref{fig:an-star-500} for different values of $x_F$.

%%%%%%%%%%%%%% Fig. 4

{}From these results we can conclude that the Sivers effect alone might in principle be able to explain the BRAHMS charged pion results on $A_N$ in the full kinematical range so far explored, as well as almost the full amount of STAR $\pi^0$ data on $A_N$. This is to be contrasted with the analogous study of the Collins effect~\cite{Anselmino:2012rq}, with the conclusion that such effect alone cannot explain the observed values of $A_N$ in the medium-large $x_F$ region.

This can be understood as follows. In the case of SSAs for neutral pion production the Collins effect suffers from two possible cancellations: the opposite sign between the $u$ and $d$ quark transversity distributions and the opposite sign between the favoured and disfavoured Collins FFs (necessary
to build the Collins FF for $\pi^0$); instead, for the Sivers effect only a cancellation between $u$ and $d$ flavours in the distribution sector may play a role, as it couples to the unpolarised TMD-FF.

A further remark concerns the values of the $\beta$ parameters and the
area spanned by the bands: the upper borderlines of the scan bands for
neutral and positively charged pions correspond to the set of Sivers functions
with $\beta_u=0$ (up quark unsuppressed) and $\beta_d=4$ (down quark strongly
suppressed), while the lower borderlines correspond to the case where the values of
$\beta$ are interchanged. Notice that larger values of $\beta$ would not change
this picture. For negative pions the situation is just reversed since to get the
largest values, in size, of $A_N$ (lower border) the down quark should dominate
(that is $\beta_d=0$ and $\beta_u=4$).

The results obtained with a different choice of the fragmentation functions (the DSS set) are qualitatively very similar in the large $x_F$ regions. They are instead smaller in size at smaller $x_F$, due to the large gluon contribution in the leading order (LO) DSS fragmentation functions. They
are not shown here.

\begin{figure}[t]
\includegraphics[width=14cm,angle=0]{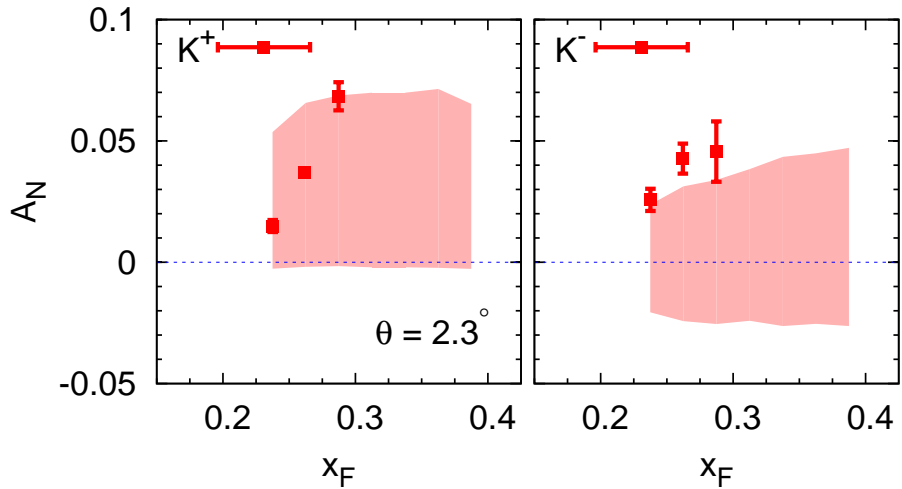}
\caption{
Scan band ({\it i.e.} the envelope of the 81 curves obtained with the scanning procedure) for the Sivers contribution to the kaon single spin asymmetry $A_N$, as a function of $x_F$, at $\sqrt s =$ 200 GeV and a fixed scattering angle, compared with the corresponding BRAHMS experimental data~\cite{Lee:2007zzh}. The shaded scan band is generated, adopting the GRV98 set of collinear PDFs and the Kretzer FF set, following the procedure explained in the text.}
\label{fig:an-brahms-kappa}
\end{figure}

The case of SSAs for kaon production would require a further study of the
corresponding unpolarised fragmentation functions, which represents an open issue
by itself and falls outside the purposes of this paper. However, for completeness
and a qualitative estimate, we consider $A_N$ for $K^\pm$ production as measured
by the BRAHMS Collaboration~\cite{Lee:2007zzh}, with the kaon set of fragmentation
function as given in Ref.~\cite{Kretzer:2000yf}. Our results for the scan band,
compared with the data, are shown in Fig.~\ref{fig:an-brahms-kappa}.

%%%%%%%%%%%%%% Fig. 5

The results in Figs.~\ref{fig:an-brahms-pi}-\ref{fig:an-brahms-kappa} show that the Sivers effect alone, as computed in our GPM scheme, Eqs.~(\ref{an})--(\ref{numans}), and based on the Sivers functions extracted from SIDIS data and assumed to be universal, can be large enough to explain alone the pion SSAs $A_N$ observed at RHIC. One should not forget that, indeed, the phenomenology of the Sivers effect was originally generated in the attempt to explain the large values of $A_N$ observed by the E704 Collaboration~\cite{Sivers:1989cc,Sivers:1990fh,Anselmino:1994tv}. However, the amount of uncertainty in the scan bands, due to lack of precise SIDIS data at large $x$, is still much too large to draw any definite conclusions.

A full understanding of the SSAs in inclusive $\pup p \to h \, X$ processes should also take into account the contribution of the Collins effect, which might be small, but not entirely negligible. Rather than addressing the issue of a best fit of SIDIS + $A_N$ data with Collins and Sivers effects, which is premature at this stage, we now adopt a more pragmatic attitude. We wonder whether, among the 81 sets of parameters which build up the possible results on $A_N$ contained in the scan bands, we can find some which give a good description of all the data.

\subsubsection{Results with a selected set of parameters and its statistical uncertainty bands}
\begin{figure}
\includegraphics[width=14cm,angle=0]{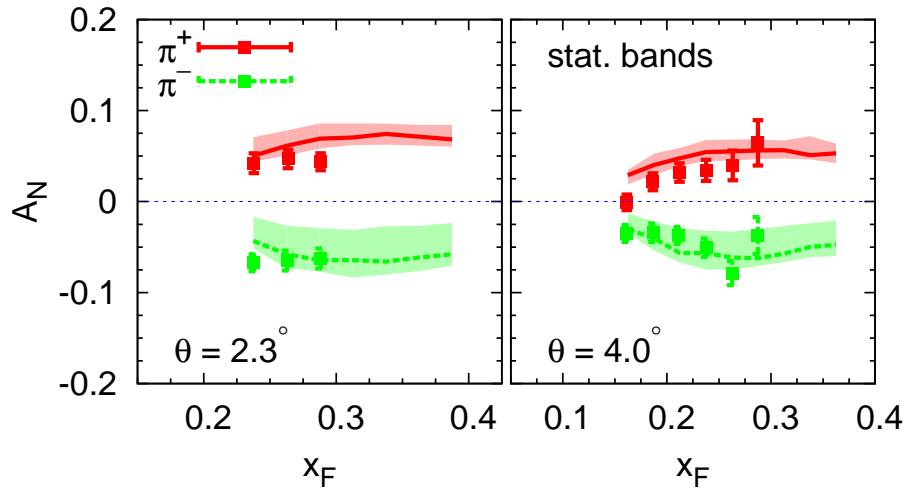}
\caption{
  The Sivers contribution to the charged pion single spin asymmetry $A_N$, compared with the corresponding BRAHMS experimental data at two fixed scattering angles and $\sqrt s = 200$ GeV~\cite{Lee:2007zzh}. The central lines are obtained adopting the GRV98 set of collinear PDFs and the Kretzer FFs, with the Sivers functions as in Eqs.~(\ref{eq:siv-par})--(\ref{eq:h-siv}) with the parameters given in Table 1. The shaded statistical error bands are generated applying the error estimate procedure described in Appendix A of Ref.~\cite{Anselmino:2008sga}.}
  \label{fig:an-brahms-band4}
\end{figure}

\begin{figure}
\includegraphics[width=14cm,angle=0]{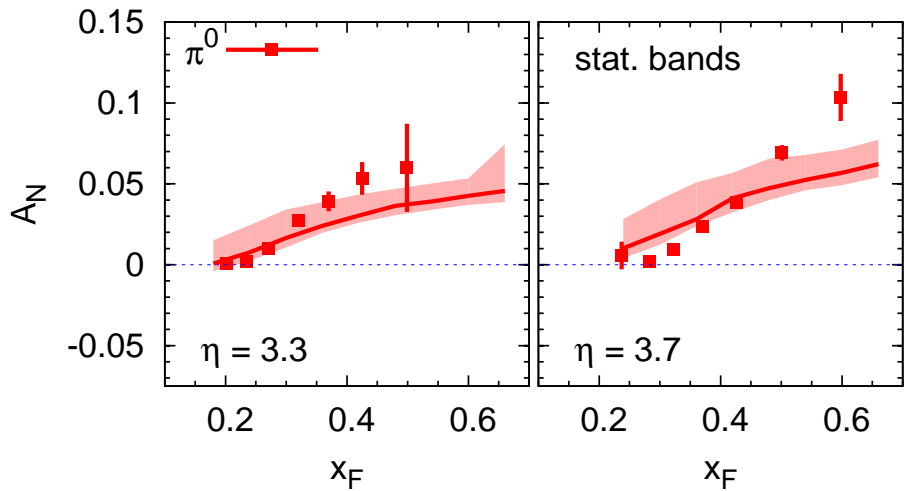}
\caption{
  The Sivers contribution to the neutral pion single spin asymmetry $A_N$, compared with the corresponding STAR experimental data at two fixed pion rapidities and $\sqrt s = 200$ GeV~\cite{Abelev:2008af}. The central lines are obtained adopting the GRV98 set of collinear PDFs and the Kretzer FFs, with the Sivers functions as in Eqs.~(\ref{eq:siv-par})--(\ref{eq:h-siv}) with the parameters given in Table 1. The shaded statistical error bands are generated applying the error estimate procedure described in Appendix A of Ref.~\cite{Anselmino:2008sga}.}
  \label{fig:an-star-band4}
\end{figure}

\begin{figure}
\includegraphics[width=14cm,angle=0]{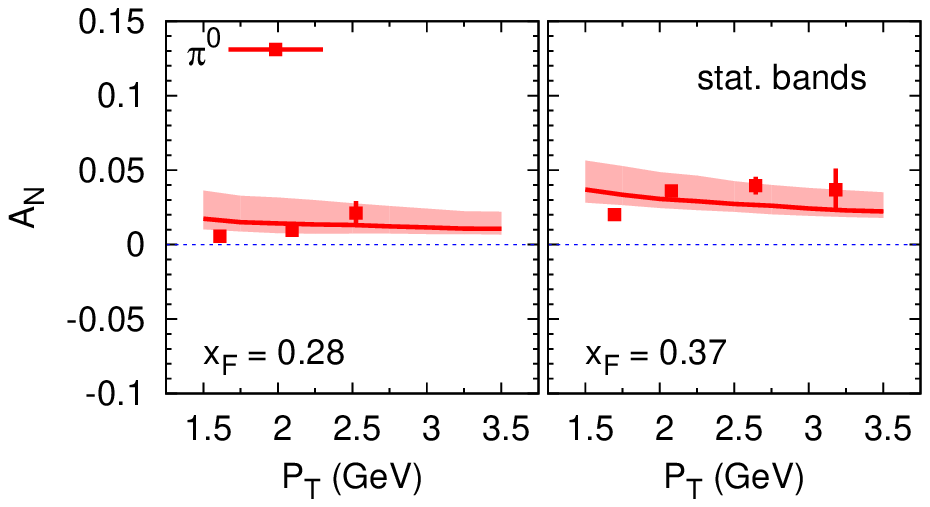}
\includegraphics[width=14cm,angle=0]{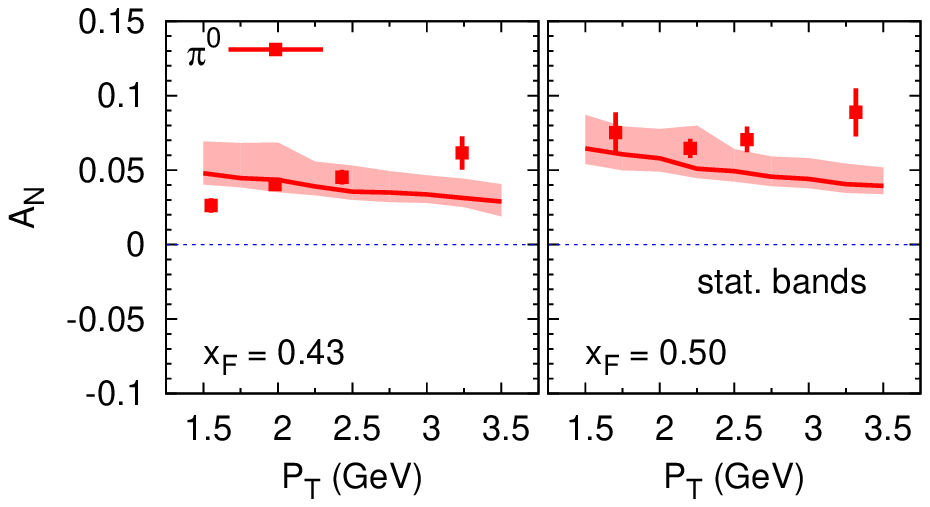}
\caption{
  The same as in Fig.~\ref{fig:an-star-band4}, but with the STAR data plotted vs.~the pion transverse momentum, $P_T$, for different bins in $x_F$, $\langle x_F \rangle$ = 0.28, 0.37, 0.43 and 0.50. } \label{fig:an-star-band4-pT}
\end{figure}

\begin{figure}
\includegraphics[width=14cm,angle=0]{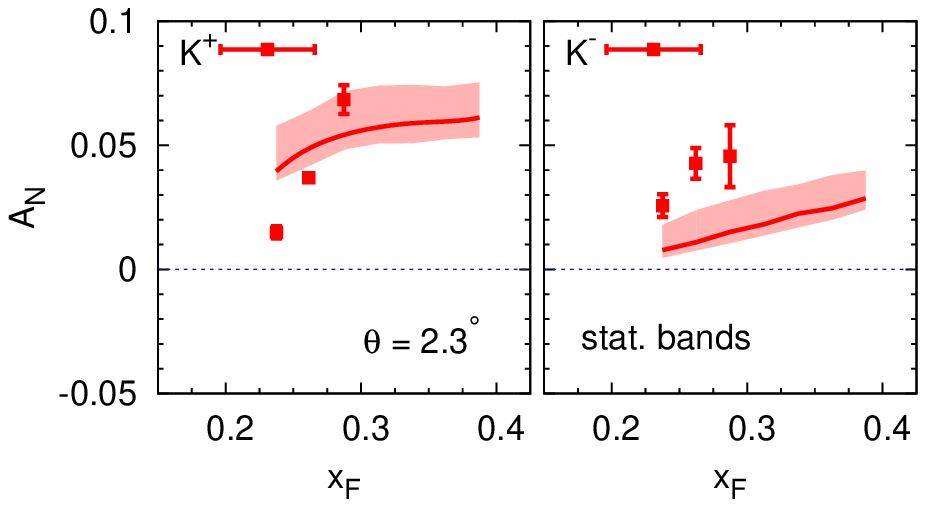}
\vskip -12pt
\caption{
  The Sivers contribution to the charged kaon single spin asymmetry $A_N$, compared with the corresponding BRAHMS experimental data at a fixed scattering angle and $\sqrt s = 200$ GeV~\cite{Lee:2007zzh}. The central lines are obtained adopting the GRV98 set of collinear PDFs and the Kretzer FFs, with the Sivers functions as in Eqs.~(\ref{eq:siv-par})--(\ref{eq:h-siv}) with the parameters given in Table 1. The shaded statistical error bands are generated applying the error estimate procedure described in Appendix A of Ref.~\cite{Anselmino:2008sga}.} \label{fig:an-brahms-kaon-band4}
\end{figure}

Among the full set of curves produced by the scan procedure, we have isolated the set leading to the best description of $A_N$ (actually one could find more than a single set); we have then evaluated, as in Appendix A of Ref.~\cite{Anselmino:2008sga}, the corresponding statistical error band. Our results are presented in Figs.~\ref{fig:an-brahms-band4}, \ref{fig:an-star-band4}, \ref{fig:an-star-band4-pT} and \ref{fig:an-brahms-kaon-band4}, respectively for BRAHMS $\pi^\pm$ data vs.~$x_F$ at fixed angles, for STAR $\pi^0$ results vs.~$x_F$ at fixed pseudo-rapidities, for STAR  $\pi^0$ results vs.~$P_T$ at different $x_F$ values, and for BRAHMS $K^\pm$ data vs.~$x_F$ at a fixed angle, all of them at $\sqrt s = 200$ GeV. The corresponding values of the parameters are given in Table 1. From these results one can see that it is possible to find a set of Sivers functions for $u$ and $d$ quarks which, while describing well the SIDIS data, can also describe fairly well, alone, the SSAs for pion production, as measured both by BRAHMS and STAR Collaborations at 200 GeV.
%
%%%%%%%%%%%%%% Fig. 6
%
%%%%%%%%%%%%%% Fig. 7
%

%%%%%%%%%%%%%% Fig. 8

\begin{figure}
\includegraphics[width=7.5cm,angle=0]{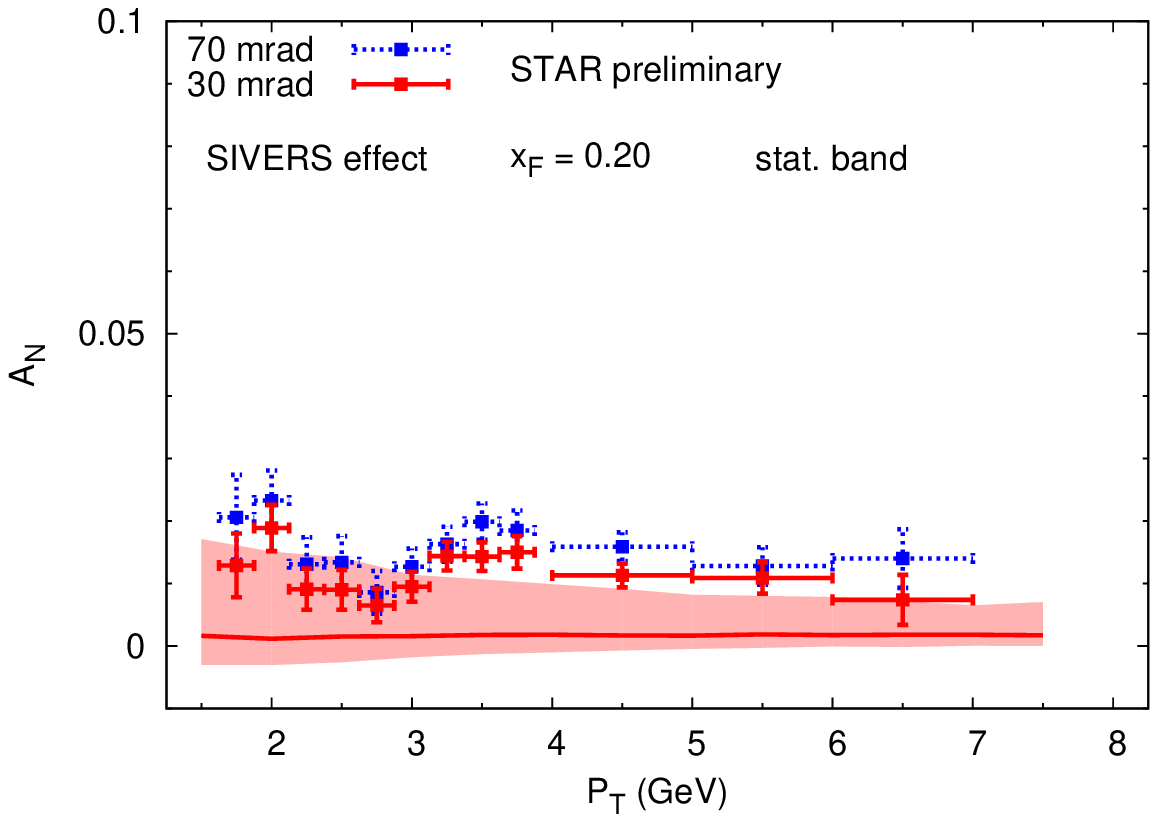}
\includegraphics[width=7.5cm,angle=0]{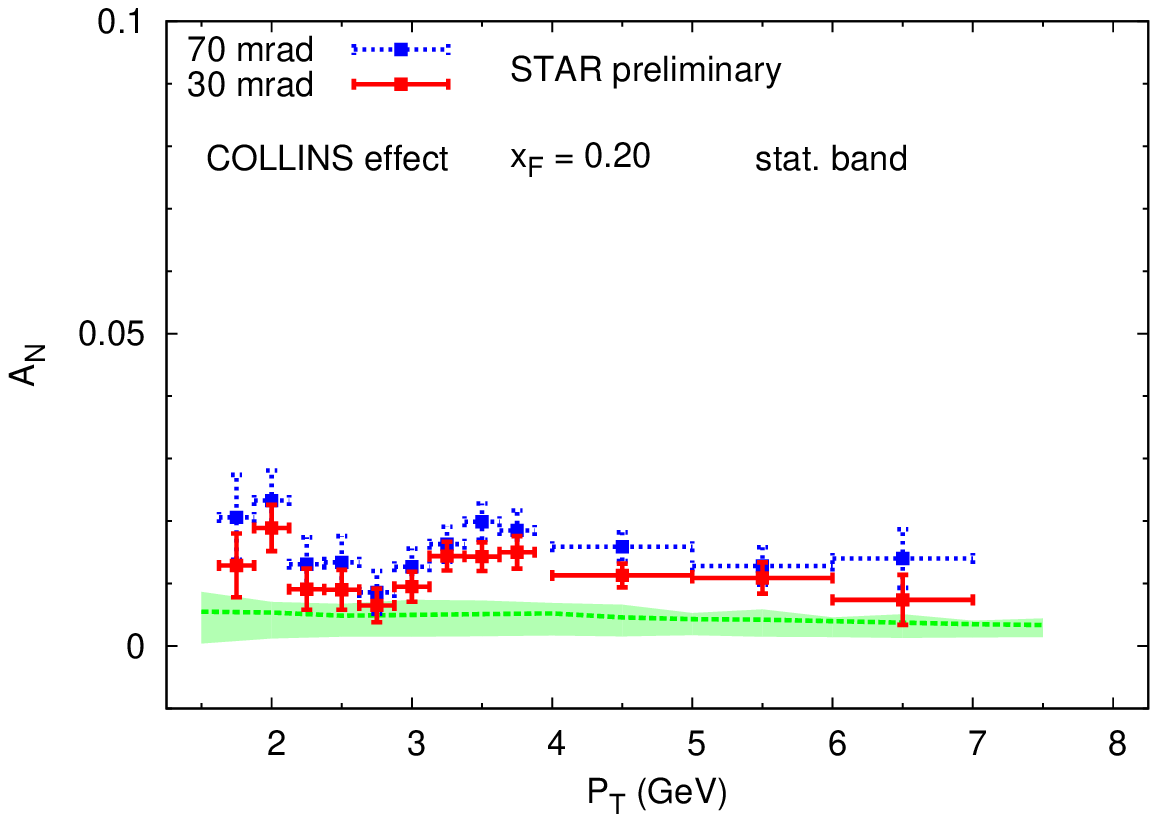}\\
\includegraphics[width=7.5cm,angle=0]{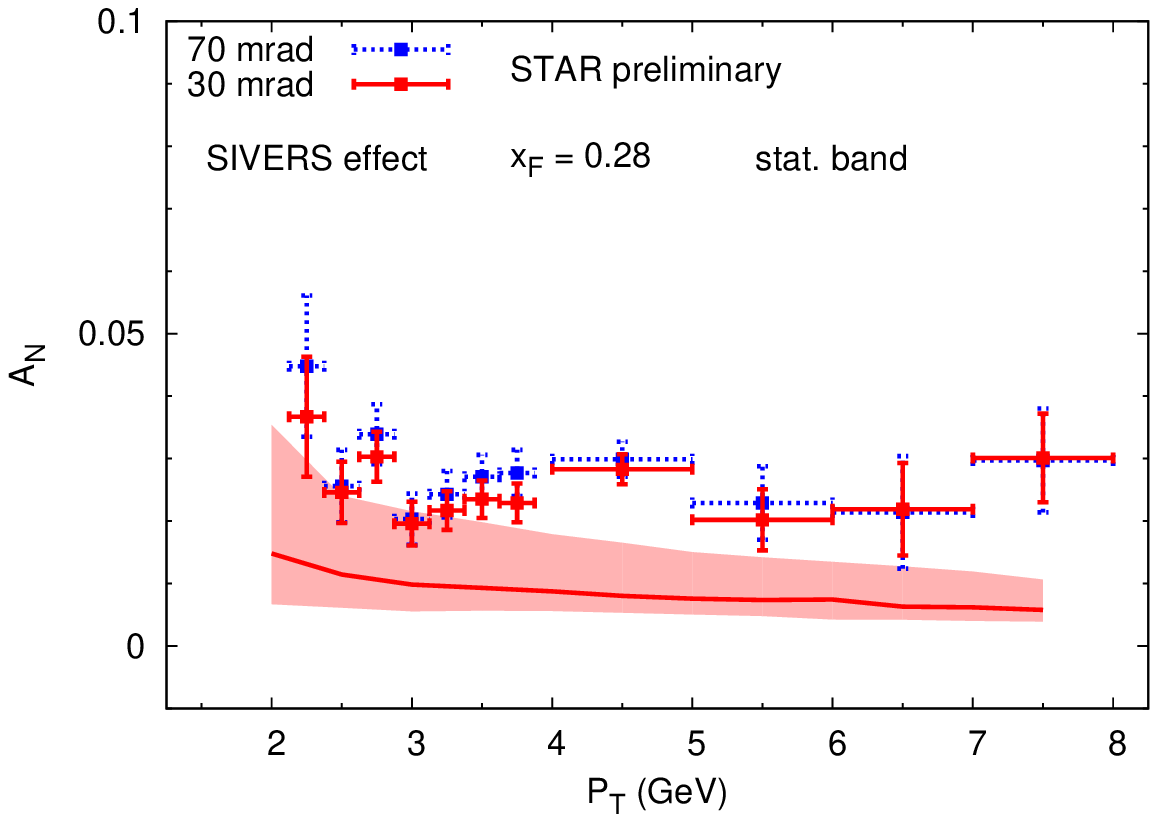}
\includegraphics[width=7.5cm,angle=0]{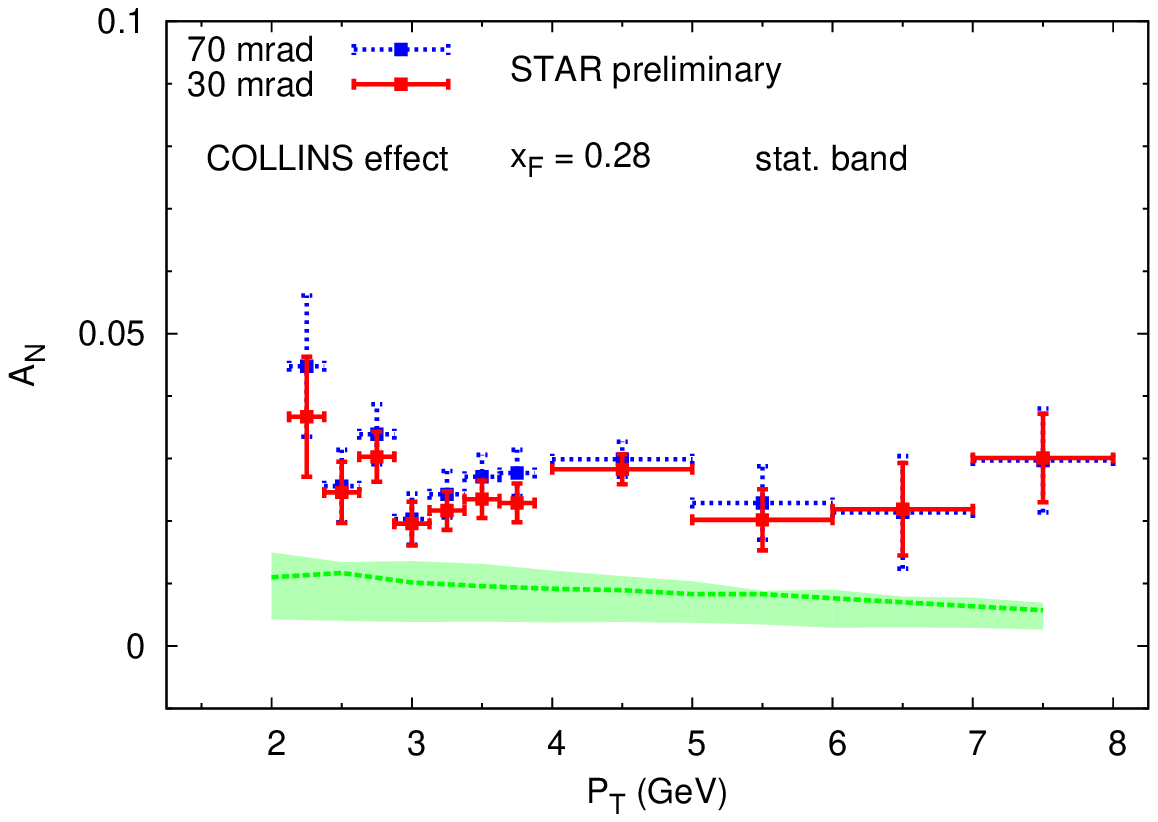}
\caption{
  {\it Left panels}: the Sivers contribution to the $\pi^0$ single spin asymmetry $A_N$ vs.~the pion transverse momentum $P_T$, for different bins in $x_F$, compared with the corresponding STAR preliminary data at $\sqrt s = 500$ GeV and $\langle x_F \rangle$ = 0.20, 0.28~\cite{Igo:2012}. The central lines are obtained adopting the GRV98 set of collinear PDFs and the Kretzer FFs, with the Sivers functions as in Eqs.~(\ref{eq:siv-par})--(\ref{eq:h-siv}) with the parameters given in Table 1. The shaded statistical error bands are generated applying the error estimate procedure described in Appendix A of Ref.~\cite{Anselmino:2008sga}.
  {\it Right panels}: the Collins contribution to the same $A_N$, computed according to Ref.~\cite{Anselmino:2012rq}, choosing the Collins functions, among those of the scan band, which give the maximum contribution.
}
\label{fig:an-star500-band4-SC}
\end{figure}

The preliminary STAR data at 500 GeV~\cite{Igo:2012} deserve a dedicated comment. Quite surprisingly, they show values of $A_N$ of the order of few percents, with a flat behaviour as a function of $P_T$ at fixed $x_F$, up to $P_T \simeq 7$ GeV. Such a trend is well reproduced by our set of chosen best parameters; however, the computed magnitude of $A_N$ is smaller than data, as shown in Fig.~\ref{fig:an-star500-band4-SC}, left plots. As the asymmetry is so small, we have also computed the Collins contribution to $A_N$, following Ref.~\cite{Anselmino:2012rq}. It turns out that, for some sets of the parameters, the Collins contribution has a similar trend and magnitude as the Sivers one, as shown in Fig.~\ref{fig:an-star500-band4-SC}, right plots. Then, an appropriate sum of the two contributions, according to Eq.~(\ref{ansc}), might well explain also this new puzzling data.

Another cautious comment about the STAR data on $A_N$ at 500 GeV concerns the large value of their QCD scale, $Q^2 = P_T^2$. As we noticed for the COMPASS proton data, at such values the TMD evolution might play an important role. Our results should then be taken as an indication in favour of a combined Collins + Sivers effect, rather than a proof. Qualitatively, one expects from TMD evolution an increase of the average $\langle \kt^2 \rangle$ value of the Sivers distribution, which would help increasing the corresponding value of $A_N$.

%%%%%%%%%%%%%% Fig. 9
%
%%%%%%%%%%%%%% Fig. 10
%

%%%%%%%%%%%%%%%%%%%%%%%%%%%% Tabella 1
\begin{table}[b]
\caption{
  Our chosen set of 7 parameters, Eq.~(\ref{eq:7-par}), fixing the $u$ and $d$ quark Sivers distribution functions, according to Eqs.~(\ref{eq:siv-par}-\ref{eq:h-siv}). Among the 81 sets of the scan procedure, this set gives the best description of the $A_N$ data. The corresponding total value of $\chi^2$ for the 217 SIDIS data points is 273.2, which is very close to the best value $\chi^2_0 = 270.5$ of the reference set. The statistical errors quoted for each free parameter correspond to the shaded uncertainty areas in Figs.~6–-9 and 11 and the left panels of Fig.~10, as explained in the text and in the Appendix of Ref.~\cite{Anselmino:2008sga}.}
\vskip 10pt
\renewcommand{\tabcolsep}{0.4pc} % enlarge column spacing
\renewcommand{\arraystretch}{1.2} % enlarge line spacing
\begin{tabular}{@{}lll}
 \hline
 $N_{u} =  0.35^{+0.08}_{-0.04}$ & $\> \> \alpha_u = 0.00^{+0.06}_{-0.00}$ &
 $ \> \> \> \> \> \> \beta_u$  = 0.00 \\
 \hline
 $N_{d} = -1.00^{+0.24}_{-0.00}$ & $\> \> \alpha_d = 0.24^{+0.11}_{-0.17}$ & $ \> \> \> \> \> \> \beta_d$  = 1.00  \\
 \hline
 $M^2 = 0.44^{+0.78}_{-0.15}$ GeV$^2$ &\\
 \hline
\end{tabular}
\end{table}
%%%%%%%%%%%%%%%%%%%%%%%%%%%%%%%%%%%%%%%%%%%%%%%%%%%%%%%%%%%%%%%%%%%%%%%%%%

\subsection{SSAs for $\pup p\to {\rm jet}\, X$ and $\pup p\to \gamma \, X$ processes}

In these processes no fragmentation mechanism is required, so that, within the GPM and the TMD factorisation approach, one can access directly the spin and $\bfk_\perp$ properties of the partonic distributions. After integration over the intrinsic azimuthal phases only the Sivers effect survives, which is then best studied in these processes, as discussed, {\it e.g.}, in Refs.~\cite{D'Alesio:2004up,D'Alesio:2010am}. Notice that, for the same reasons, the SSAs for inclusive jet or photon production can be used to test the process dependence of the Sivers functions in a modified generalised parton model with inclusion of initial and final state interactions~\cite{Gamberg:2010tj,D'Alesio:2011mc} or within the twist-3 approach~\cite{Gamberg:2013kla}.

The numerator of $A_N$ for the inclusive jet production can be obtained from Eq.~(\ref{numans}) simply replacing the TMD fragmentation function, $D_{h/c}(z,p_\perp)$, with a factor $\delta(z-1)\,\delta^2(\bm{p}_\perp)$ (and identifying now the final hadron momentum, $\bfp_h$, with the jet momentum $\bfp_c \equiv \bfp_{\rm jet}$). More explicitly the numerator of $A_N$ for inclusive jet production reads
\bea
[d\sigma^\uparrow - d\sigma^\downarrow]_{\rm Sivers}^{\pup p\to {\rm jet}\;X}
&=& \!\!\! \sum_{a,b,c,d} \int \frac{dx_a \, dx_b}
{16 \, \pi^2 \, x_a \, x_b \, s} \; d^2 \bfk_{\perp a} \, d^2 \bfk_{\perp b}\,
\delta(\hat s + \hat t + \hat u) \nonumber\\
&\times& \Delta^N\!f_{a/\pup}(x_a, k_{\perp a}) \,
\cos (\phi_a)\,
 f_{b/p}(x_b, k_{\perp b}) \> \frac{1}{2}
 \left[ |\hat M_1^0|^2 + |\hat M_2^0|^2 + |\hat M_3^0|^2 \right]_{ab\to cd} \>
 \>. \label{numans-jet}
\eea
Notice that the elementary hard scattering interactions are exactly the same as those for the inclusive hadron production and the jet, at LO, is identified with the final parton $c$.

Concerning the direct photon production the basic partonic processes are the Compton process $g\,q\,(\bar q) \to \gamma \, q\,(\bar q)$ and the annihilation process $q \, \bar q\to \gamma \, g$. In this case one can formally use the above equation replacing the partonic unpolarised cross section, Eq.~(\ref{eq:sigma}), with the corresponding one for the process $a \, b \to \gamma \, d$ (see also Ref.~\cite{D'Alesio:2004up}).

%%%%%%%%%%%%%% Fig. 11
\begin{figure}[hb]
\includegraphics[width=7cm,angle=0]{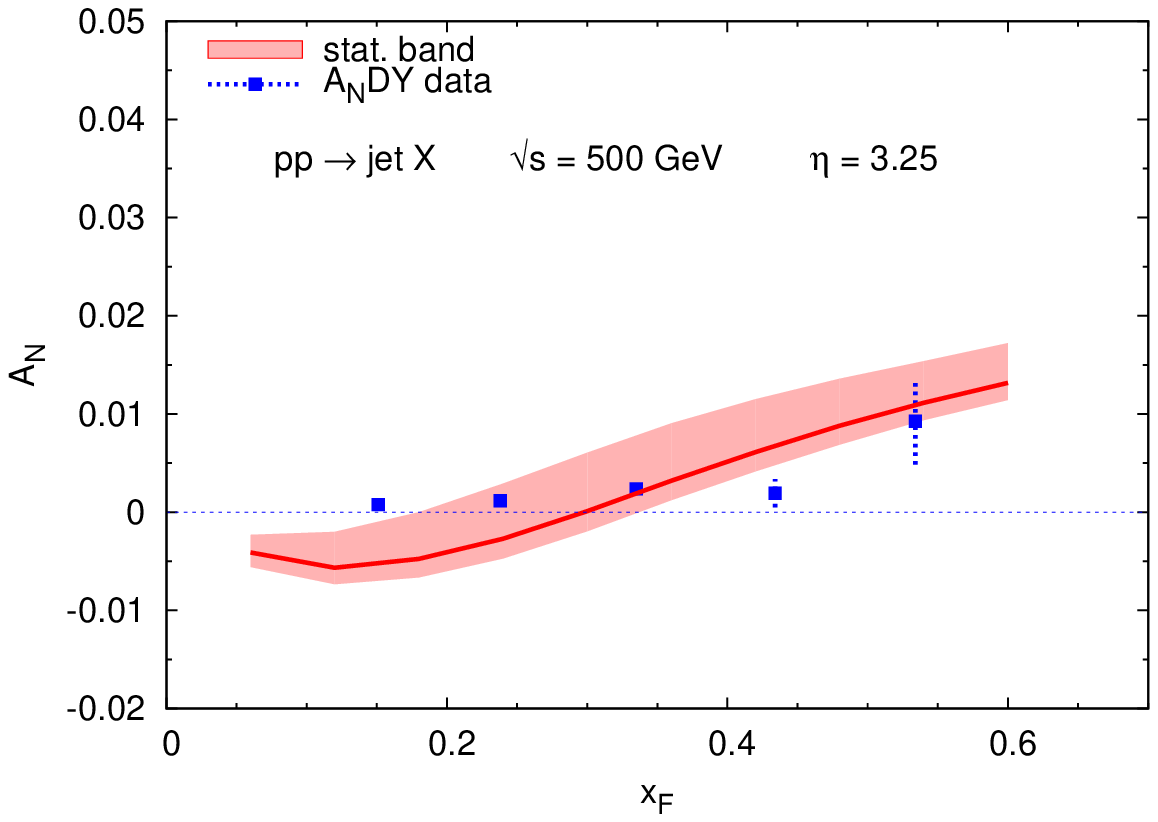}
\includegraphics[width=7cm,angle=0]{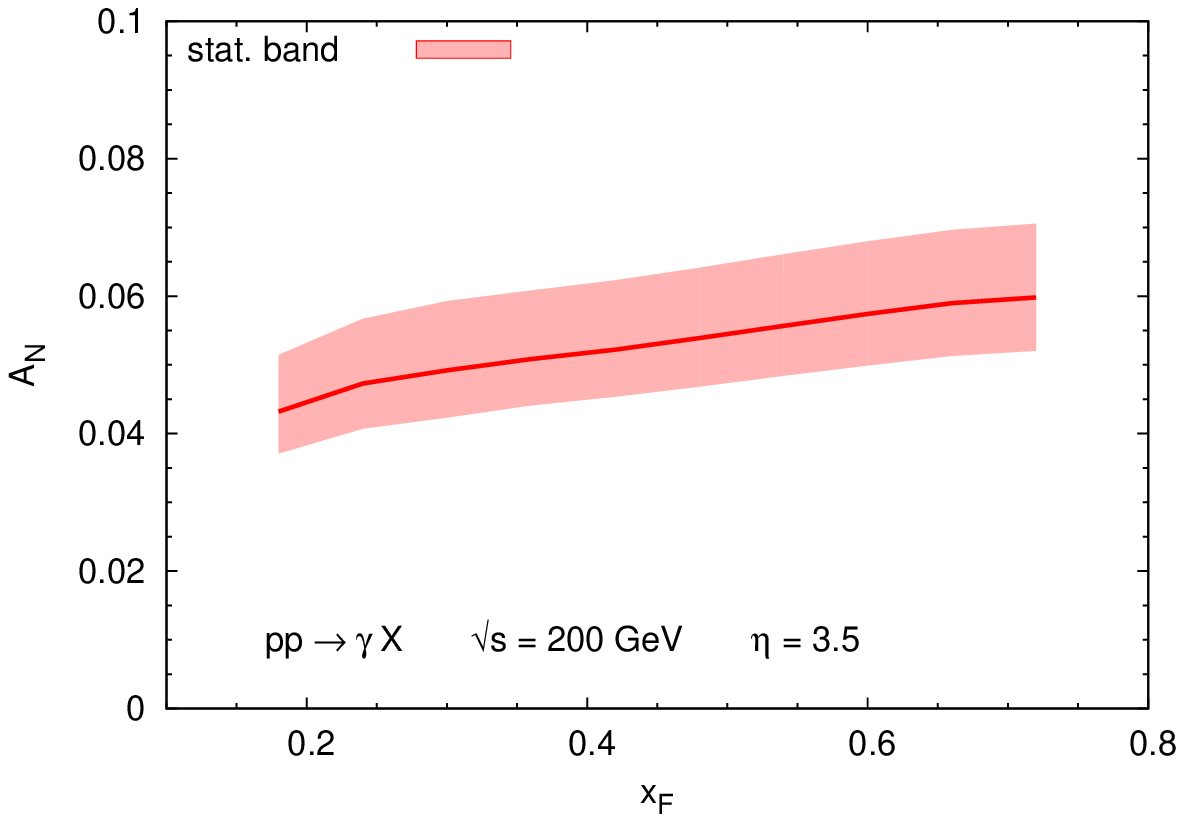}
\caption{
  {\it Left panel}: our estimate for the jet SSA $A_N$ at $\sqrt s$ = 500 GeV, as a function of $x_F$ at fixed pseudo-rapidity $\eta= 3.25$, compared with the A$_N$DY data~\cite{Bland:2013pkt}. The central line is obtained adopting the GRV98 set of collinear PDFs, with the Sivers functions as in Eqs.~(\ref{eq:siv-par})--(\ref{eq:h-siv}) with the parameters given in Table 1. The shaded statistical error band is generated applying the error estimate procedure described in Appendix A of Ref.~\cite{Anselmino:2008sga}.
  {\it Right panel}: the same estimate as in the left panel for a direct photon, rather than a jet, production at $\sqrt s$ = 200 GeV and $\eta=3.5$. }
  \label{fig:an-andy-gamma}
\end{figure}
%
%%%%%%%%%%%%%% Fig. 12
\begin{figure}[b]
\includegraphics[width=9cm,angle=0]{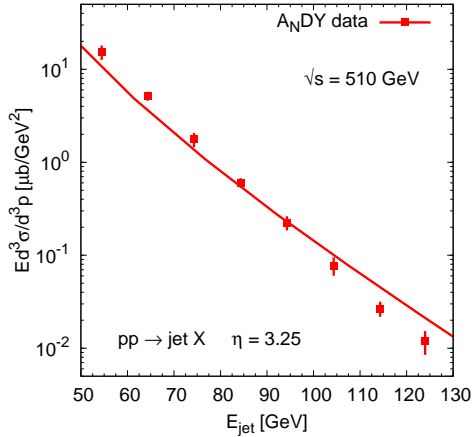}
\caption{
  Our computation of the unpolarised cross section for jet production vs.~the jet energy, at $\sqrt s$ = 510 GeV and fixed pseudo-rapidity $\eta = 3.25$, compared with A$_N$DY data~\cite{Bland:2013pkt}.}
  \label{fig:unp-andy-jet}
\end{figure}

No SSA data are so far available for direct photon production, while very recently some preliminary data for inclusive jet production have been released by the A$_N$DY Collaboration at $\sqrt s = 500$
GeV~\cite{Bland:2013pkt}. The values measured for $A_N$ are very tiny, but very precise and might indicate a non zero asymmetry.

In the left plot of Fig.~\ref{fig:an-andy-gamma} we show our estimate, based on the chosen best set parameters of Table 1, for $A_N(x_F)$ in $\pup p\to {\rm jet}\, X$ processes at a fixed pseudo-rapidity value and $\sqrt s$ = 500 GeV, and compare it with the A$_N$DY data~\cite{Bland:2013pkt}.
In the right plot we give our corresponding estimates for $A_N(x_F)$ in $\pup p \to \gamma \, X$ processes at a fixed pseudo-rapidity value and $\sqrt s$ = 200 GeV.

For consistency, in Fig.~\ref{fig:unp-andy-jet} we compare our (leading order) computation of the cross section for jet production as given by Eq.~(\ref{numans-jet}) where we replace the factor $\Delta^N\!f_{a/\pup}\,\cos(\phi_a)$ with $f_{a/p}$, with the A$_N$DY data at $\sqrt s = 510$ GeV and fixed pseudo-rapidity $\eta = 3.25$.

\section{Conclusions}

The origin of the azimuthal asymmetries in SIDIS processes is considered to be
well understood and related to TMD-PDFs and TMD-FFs, via the QCD TMD factorisation
scheme. Indeed, the measurement of such asymmetries has been used to extract
information on the TMDs. A reasonable knowledge of the Sivers TMD-PDF is by now
available and confirmed by independent groups~\cite{Anselmino:2005ea,Anselmino:2008sga,Vogelsang:2005cs,Collins:2005ie, Anselmino:2005an}. However, due to the kinematical range of the data, this
knowledge is limited to the small $x$ region, $x \lsim 0.3$. The same TMD factorisation is expected to hold also in Drell-Yan and
$e^+e^- \to h_1 \, h_2 \, X$ processes.

The situation is not so clear concerning the oldest and largest single spin
asymmetries $A_N$ measured in several hadronic processes, in particular in
$\pup p \to h \, X$. Due to the presence of a single large scale -- the $P_T$ of
the final hadron -- one cannot extend to these processes the proof of the QCD
TMD factorisation theorem, which requires the presence of two separate scales,
a large and a small one. As explained in the Introduction, a twist-3 collinear
and factorised approach has been proposed~\cite{Efremov:1981sh,Efremov:1984ip,
Qiu:1991pp,Qiu:1998ia,Kanazawa:2000hz,Kanazawa:2000kp,Kouvaris:2006zy,Kanazawa:2010au,Kanazawa:2011bg}, which introduces new three-parton correlation functions,
related to the $k_\perp$-moment of the TMD-PDFs. However, it seems to predict
values of $A_N$ opposite to the observed ones~\cite{Kang:2011hk}.
Thus, the true origin of the large values of $A_N$ remains obscure.

Among the first attempts to explain $A_N$~\cite{Sivers:1989cc,Sivers:1990fh,
Anselmino:1994tv,Anselmino:1998yz,Anselmino:1999pw, D'Alesio:2004up,Anselmino:2005sh,D'Alesio:2007jt}, and the unpolarised cross section~\cite{Feynman:1978dt},
one should consider the simple extension of the collinear QCD factorisation to
the TMD case, the so-called Generalised Parton Model (GPM) in which one assumes
TMD factorisation and the universality of the TMF-PDFs and TMD-FFs. Although
such a factorisation has not been proven, it is worth exploring its
phenomenological consequences. In this paper we have studied, within the GPM, the contribution of the Sivers effect to the single spin asymmetry $A_N$ as measured
by RHIC Collaboration experiments. The Sivers functions are the same as those
which explain azimuthal asymmetries in SIDIS processes. A similar analysis was performed in Ref.~\cite{Anselmino:2012rq}, concerning the Collins effect.

Our results, limited to the contribution of the valence quarks, to the SIDIS cases where the TMD evolution is not expected to be relevant and to the $p \, p \to \pi \, X$ large $P_T$ processes for which we can well reproduce the unpolarised cross sections, are rather encouraging. For most pion data the Sivers effect alone could explain the observed values of $A_N$ in magnitude and, in particular, in sign. This is in contrast to other approaches, also related to the Sivers effect, which seem to have
severe problems~\cite{Kang:2011hk,Gamberg:2010tj} in explaining the sign of the observed $A_N$.

We have performed our analysis by varying the parameters of the Sivers functions
which are not well fixed by the SIDIS data, due to their limited kinematical range, obtaining the so-called scan bands. In particular, we have let the power $\beta$
which fixes the large $x$ behaviour of the $u$ and $d$ quark Sivers functions,
$(1-x)^\beta$, vary between 0 and 4 in steps of 0.5. Thus, we have 81 different
sets of Sivers functions; each of them still fits well the SIDIS data. The bands, which appear in Figs. 1-5, are the envelope of the 81 different curves, $A_N(x_F)$
and $A_N(P_T)$, obtained in our GPM approach.

Then, among the explored sets of parameters, we have chosen a particular one, which gives one of the best descriptions of the $A_N$ data. We have used such a set to compute estimates for SSAs in $\pup p \to {\rm jet} \, X$ and $\pup p \to \gamma \, X$ processes. Such measurements will further allow one to discriminate between our approach and others. Predictions for $A_N$ in different processes and kinematical regions can be easily obtained, if necessary.

While encouraged by the results of our analysis we should avoid making definite conclusions at this stage. This work shows that the GPM TMD factorisation scheme could explain at the same time the main features of the SSAs measured in SIDIS and hadronic processes. While such a scheme is well justified for SIDIS it can, so far, only be considered as a phenomenological model for hadronic processes, which needs further confirmation or disproval from data and further theoretical work.
Our choice of the sets of parameters given in Table 1 is not meant to be interpreted as the final best set of Sivers functions. A full analysis, including TMD evolution, of all data involving the Sivers
effect -- {\it i.e.} all the SSAs in several different processes -- would require much more attention and work.

\acknowledgments
We would like to thank L.~Bland for information on the A$_N$DY data and S.~Heppelmann for information on the STAR data on $A_N$ at 500 GeV. We acknowledge support of the European Community under the FP7 ``Capacities - Research Infrastructures'' program (HadronPhysics3, Grant Agreement 283286). Some of us (M.A., M.B., U.D., F.M.) acknowledge partial support from MIUR under Cofinanziamento PRIN 2008. U.D.~is grateful to the Department of Theoretical Physics II of the Universidad Complutense of Madrid for the kind hospitality extended to him during the completion of this work.

\end{document}